\documentclass[fleqn,twoside]{article}
\usepackage{espcrc2}

\usepackage{graphicx}
\usepackage[figuresright]{rotating}

\newcommand{\AmS}{{\protect\the\textfont2
  A\kern-.1667em\lower.5ex\hbox{M}\kern-.125emS}}

\newcommand \VEV [1] {\left\langle{#1}\right\rangle}

\newcommand{\beq}{\begin{eqnarray}}
\newcommand{\eeq}{\end{eqnarray}}

\newcommand{\LMS}{\Lambda_{\overline{\rm MS}}}
\newcommand{\gev}{\rm GeV}
\def\jhep#1#2#3{J. High Energy Phys. {\bf #1} (#2) #3}
\def\prd#1#2#3{Phys.\ Rev.\ {\bf D#1} (#2) #3}
\def\npb#1#2#3{Nucl.\ Phys.\ {\bf B#1} (#2) #3}
\def\plb#1#2#3{Phys.\ Lett.\ {\bf B#1} (#2) #3}

\title{O.P.E. and Power Corrections to the QCD coupling constant.}

\author{Ph. Boucaud\address[LPT]{Laboratoire de Physique Th\'eorique, 
	Universit\'e de Paris XI,\\ Batiment 210, 91405 Orsay-Cedex, France.}, 
	F. De Soto\address[US]{Dpto. de F\'{\i}sica At\'omica Molecular y 
	Nuclear, Universidad de Sevilla.\\ Avda. Reina Mercedes s/n, 41012 
	Sevilla, Spain.}, A. Donini\addressmark[LPT], J.P. Leroy\addressmark[LPT], 
	A. Le Yaouanc\addressmark[LPT], J. Micheli\addressmark[LPT], 
	H. Moutarde\addressmark[LPT], O. P\`ene\addressmark[LPT], 
	J. Rodr\'{\i}guez-Quintero\address[UHU]{Dpto. de F.A., E.P.S. La R\'abida, 
	Universidad de Huelva, \\ 21819 Palos de la Fra., Spain.}}
      
\begin{document}

\begin{abstract}
Lattice data seems to show that power corrections should be convoked to
describe appropriately the transition of the QCD coupling constant
running from U.V. to I.R. domains. Those power corrections  for the
Landau-gauge MOM coupling constant in a pure Yang-Mills theory ($N_f=0$)
are analysed in terms   of Operator Product Expansion (O.P.E.) of two-
and three-point Green functions, the gluon condensate $<A^2>$  emerging
from this study. The semi-classical picture given by instantons can  be
also used to look for into the nature of the power corrections and
gluon condensate.
\vspace{1pc}
\end{abstract}

\maketitle

\section{The non-perturbative QCD coupling constant from lattice QCD.}

The non-perturbative calculation of the QCD coupling constant is a major
open question, as it could give some insight into the problem of
confinement. This calculation, as a very important task, has already been
performed by several different methods,\cite{alat,alpha}. The method used
here is based in the calculation of the coupling constant from the three
gluon  coupling \cite{alles}, that allows an easier physical
interpretation than the Schr\"odinger functional method, for example,
and if it does not allow to study a quite wide range of energies, it will
allow us to focus on the regime of  energies where the transition from
perturbative to non-perturbative  regimes in QCD occurs.

The calculation of the coupling constant in this method comes from the 
lattice evaluation of two- and three-point gluon Green functions, the renormalised
coupling being defined (according to MOM schemes) as the renormalised three gluon 
vertex where external propagators are explicitly amputated:

\beq
g(k^2)=\frac{G^{(3)}(p_1^2,p_2^2,p_3^2)\left( Z_3(k^2) \right)^{3/2}}
{ G^{(2)}(p_1^2) G^{(2)}(p_2^2) G^{(2)}(p_3^2) }\ ,
\eeq

\noindent where $G^{(2)}(p_1^2)$ and $G^{(3)}(p_1^2,p_2^2,p_3^2)$ are two- 
and three- point Green Function form factors \cite{alles,power}, and 
$Z_3(k^2)=k^2 G_R^{(2)}(k^2)$ is the gluon propagator renormalization constant.

\section{O.P.E. and $\VEV{A^2}$ condensate.}

Lattice calculations of the QCD coupling constant suggest the necessity 
to add power corrections to the purely perturbative expressions, to correctly 
describe its running \cite{power}. An Operator Product Expansion (O.P.E.) 
analysis of the Green functions in Landau gauge\footnote{In the lattice we 
will work in the minimum $A^2$ Landau gauge, $\partial_\mu A_\mu =0$ (Absolute 
Landau Gauge), so all gauge dependent quantities will be expressed in this particular 
gauge.} relates this power corrections to the existence of a non-perturbative 
$\VEV{A^2}$ condensate \cite{ope}, through expressions:

\beq
G^{(2)}_{O.P.E.}(p^2)=G^{(2)}_{Pert.}(p^2)+c\frac{\VEV{A^2}_{R,\mu}}{p^2},\nonumber \\
\alpha_s^{O.P.E}(p^2)=\alpha_s^{Pert.}(p^2)+c'\frac{\VEV{A^2}_{R,\mu}}{p^2},
\label{ope}
\eeq

\noindent where perturbative expressions are developed at three loops, and the
functions $c$ and $c'$ include the Wilson coefficient of the expansion and 
the anomalous dimension of the  condensate at leading logarithm.

By performing a combined fit of lattice results in a wide region of
energies (from $3$ to $10\ \gev$) to expressions in (\ref{ope}), in two
different  ${\rm MOM}$ schemes, a value of $\LMS$ is extracted, in  fairly
good agreement with the one obtained by the ALPHA collaboration
\cite{alpha}, by a completely  different method. A value of the
$\VEV{A^2}$  condensate comes out from the analysis.

The physical meaning of this condensate is  still
an open question, and a  lot of work is being devoted to its study during last
years, for example,  in relation to confinement \cite{Kondo}. The aim in
this work will be to study the possible semiclassical contribution to this
condensate coming from  instantons, and whether they might explain the presence
of power corrections in  Green Functions.

\section{The role of instantons.}

Instantons \cite{thooft} are classical solutions of QCD equations of motion 
that have been longly studied as a possible description of QCD vacuum,
and claimed to explain some non-perturbative properties of QCD at 
low energies, as  the axial anomaly, $\eta'$ mass, 
chiral condensate, etc (See \cite{shuryak} for a general overview). 

Based in the fact that instantons are saddle points of QCD action,
we could think of factorizing the path integral into an 
integral over semi-classical gauge field configurations (in
this case instantons) and the integral of quantum fluctuations
around this semi-classical background. This means that the gauge
field ($A_\mu^a$) could be decomposed into a instantonic background 
($(A_\mu^a)_{Inst.}$) plus quantum fluctuations:

\beq
A_\mu^a=(A_\mu^a)_{Inst.}+Q_\mu^a((A_\mu^a)_{Inst.})\ ,
\eeq

\noindent where in principle quantum fluctuations could deppend on the 
semi-classical background. 

If we nevertheless assume the hypothesis  that quantum fluctuations
are not sensitive to the  background \cite{a2}, the O.P.E.
$\VEV{A^2}$-condensate, corresponding to the non-perturbative part of
$\VEV{A^2}$, will be:

\beq
\VEV{A^2}_{O.P.E.}\approx \VEV{(A_\mu^a)_{Inst.}^2} \approx \VEV{A^2-Q^2(0)}\ .
\eeq

The task in this work will be to compute the instantonic contribution 
to this condensate, that (up to other non-perturbative contributions,
that could of course exist) we will finally compare to the O.P.E.
one, computed in \cite{ope}.

In order to compute $\VEV{A^2}$ in an instantonic background, we will 
consider the gauge field in an ensemble of non-interacting SU(3) 
instantons (${\rm I}$) and anti-instantons (${\overline{\rm I}}$) 
with radii $\{\rho_i\}$, centered in $\{z_i\}$, and with an
orientation in color space given by $\{(O^{a\alpha})_i\}$. If they are 
not interacting and randomly oriented, their contribution to the 
$\VEV{A^2}$ condensate in the absolute Landau gauge would give:

\beq
\VEV{A^2}_{\rm inst} \approx 
\frac{N}{V} \int d^4x A_\mu^a(x) A_\mu^a(x) = 12 \pi^2 \rho^2 n ,
\label{a2i}
\eeq

\noindent where $A_\mu^a(x)$ is the standard 't Hooft Polyakov instanton 
gauge field \cite{thooft}, $\rho$ the average radius, and 
$n=\frac{N_I+N_{\overline{\rm I}}}{V}$ the density. 

If we accept the phenomenological values assigned to $n$ and $\rho$ by the 
Instanton Liquid Model \cite{shuryak} ($n\sim 0.5 fm^{-4}$ and 
$\rho \sim 1/3 fm$), the instantonic contribution will be 
$\VEV{A^2}_{Inst.}\sim 0.5 \gev^2$. We will perform, however,
our own analysis, thus testing the latter approach.

\subsection{Cooling.}

In principle, a direct measure of $A^2$ in the lattice should be possible, but 
the presence  of the UV divergent part is hardly separable from the soft, 
instantonic one. The other possibility is to perform a cooling procedure, 
that, through a progressive elimination of UV fluctuations, 
will allow us to compute the number and size of instantons, giving 
an indirect measure of the $\VEV{A^2}$ condensate through (\ref{a2i}). Both methods agree
after a high number of cooling sweeps, when  quantum 
fluctuations are almost completely suppressed, but then cooling bias is so strong, as 
we will see, that no information can be recovered about the original situation.

We will use the traditional cooling method \cite{teper}, even if it introduces a number 
of known biases, as ${\rm I}-{\overline{\rm I}}$ annihilation, and a modification of
instanton sizes and lattice spacing. The approach proposed here is to compute 
instanton properties for different number of cooling sweeps, and extrapolate back
to the thermalised situation \cite{negele}, in order to recover their physical meaning\footnote{
The use of improved cooling methods, as the one developed in \cite{garciaperez}, could 
improve this approach, as radii evolution is minimised, but ${\rm I}-{\overline{\rm I}}$ 
annihilation is unavoidable, so the extrapolation will be anyway necessary.}.

\subsection{Shape Recognition.}

Instantons will be localised in cooled lattices via a geometrical method (Described 
in \cite{instanton}.) that accepts a topological charge lump as an instanton when the 
ratio of the integral over lattice sites with topological charge bigger than a given 
fraction, $\alpha$, of the topological charge at the maximum and its theoretical 
counterpart,

\beq
\epsilon(\alpha)=\frac{\int_{x/\frac{|Q_\rho(x)|}{|Q_\rho(0)|}\ge \alpha} d^4x Q_\rho(x)}
{1-3\alpha^{1/2}+2\alpha^{3/4}}\ ,
\label{epsilon}
\eeq

\noindent is around $1$ for a range of values of $\alpha$. To verify that it is really 
an instanton, we compute (\ref{epsilon}) both for the topologial charge density and the action,
accepting it as an instanton only if it is self-dual.

Once the lump has been identified as an instanton, the radius will be computed 
from the size of the cluster where the integral has been developed, in agreement 
with the value of the radius deduced from the value of the topological charge at 
the maximum.

We have tested this method for different lattice spacings, from 
$0.055$ to $0.23$ fm. As a matter of fact, for big lattice spacings, 
instantons have radii of order of the lattice spacing, being in practice 
indistinguishable of UV fluctuations. Therefore, only lattices finer 
than $0.1$ fm will be used, otherwise an important number of instantons with small radii
will be lost (all for radii below two lattice spacings).

\subsection{Results.}

\begin{figure}[htb]
\begin{center}
\includegraphics[width=15pc]{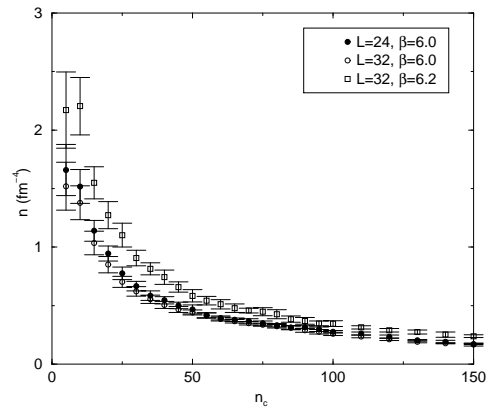}
\vspace{-0.5cm}
\caption{\small {\it Density of instantons measured form different lattice volumes 
and lattice spacings ($a(\beta=6.0)=0.101\ {\rm fm},\ a(\beta=6.2)=0.074\ {\rm fm}$)}}
\vspace{-0.5cm}
\label{comp}
\end{center}
\end{figure}

If we compare the measures of the instanton density for two different 
lattice spacings, $0.074$ and $0.101\ {\rm fm}$, and different volumes, 
as a function of the number of cooling sweeps (Figure \ref{comp}), 
we can conclude that no finite volume effect affects our results, as the 
measures for the $24^4$ lattice and $32^4$ one are compatible. This is true
excepting lattice volumes approaching the instanton volume.

For different values of the lattice spacing, on the contrary, bigger densities are found
at the same number of cooling sweeps for finer lattices. In addition to the 
problems to detect instantons with small radii, there is a known problem of 
cooling procedures, they are not independent of the lattice spacing \cite{teper}, and both
effects could play an important role to explain the difference found between different lattice 
simulations.

\begin{figure}[htb]
\begin{center}
\includegraphics[width=15pc]{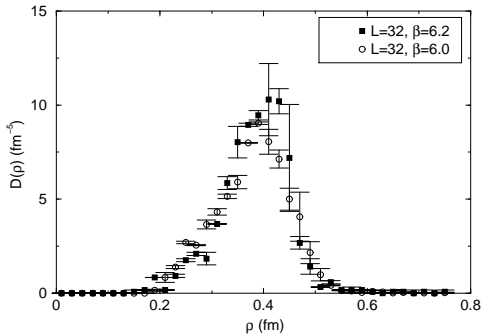}
\vspace{-0.5cm}
\caption{\small {\it Comparison on instanton histogram at $n_c=10$ 
for $\beta=6.0$ and $n_c=15$ for $\beta=6.2$.}}
\vspace{-0.5cm}
\label{hist}
\end{center}
\end{figure}

To illustrate this, if we fix the value of the density (that is, we
perform a different number of  cooling sweeps for each lattice
spacing) and compare the average radii, they agree within  the error
limits. Moreover, comparing the distribution of instanton radii, they
are in a  quite good agreement (Figure \ref{hist}), even with very
few statistics (10 configuration each).

\subsection{A naive model of annihilation.}

With the method outlined above, we compute the density and size of instantons in a lattice, 
for different numbers of cooling sweeps, $n_c$, obtaining values with a strong dependence on 
$n_c$ (See figures), that avoids to obtain any physical information at fixed $n_c$. Moreover, 
this dependence on the number of cooling sweeps is different at different lattice spacing, so
an extrapolation of results after cooling is unavoidable to obtain any physical information.

As a first approach to the understanding of this evolution, we will make a 
simple model, where instantons annihilate with antiinstantons (Being so 
$\Delta N=N_I-N_{\overline{I}}$ a constant) proportionally to their packing ratio, 
and to the number of antiinstantons, so that the equation for the evolution of 
$N=N_I+N_{\overline{I}}$ is:

\beq
\frac{\partial N}{\partial n_c}\ = - \frac{\lambda}{2V}\rho^4(n_c) (N(n_c)^2-\Delta N^2).
\label{nnc}
\eeq

\begin{figure}[ht]
\begin{center}
\begin{tabular}{c}
\includegraphics[width=16pc]{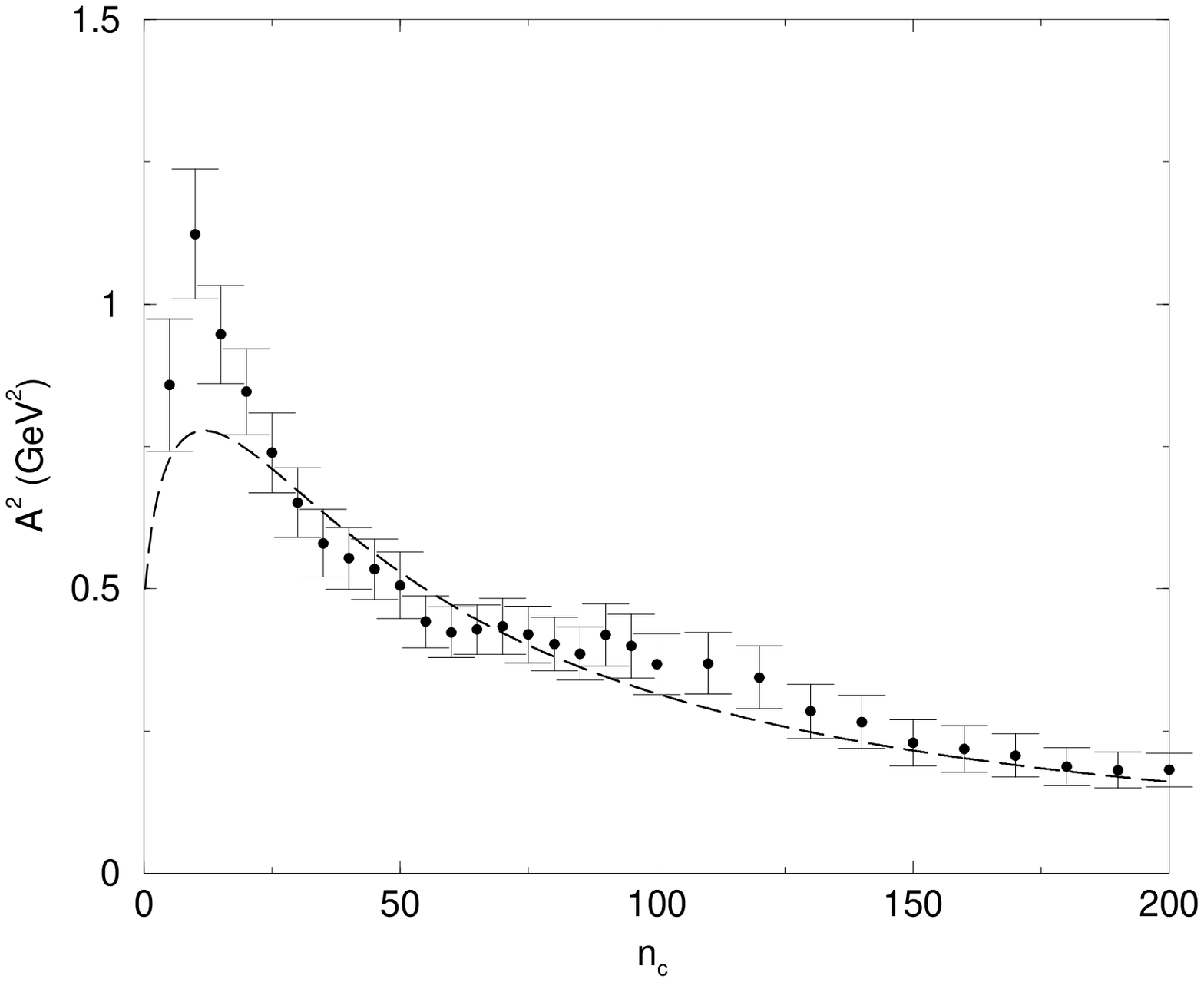}  \\
\includegraphics[width=16pc]{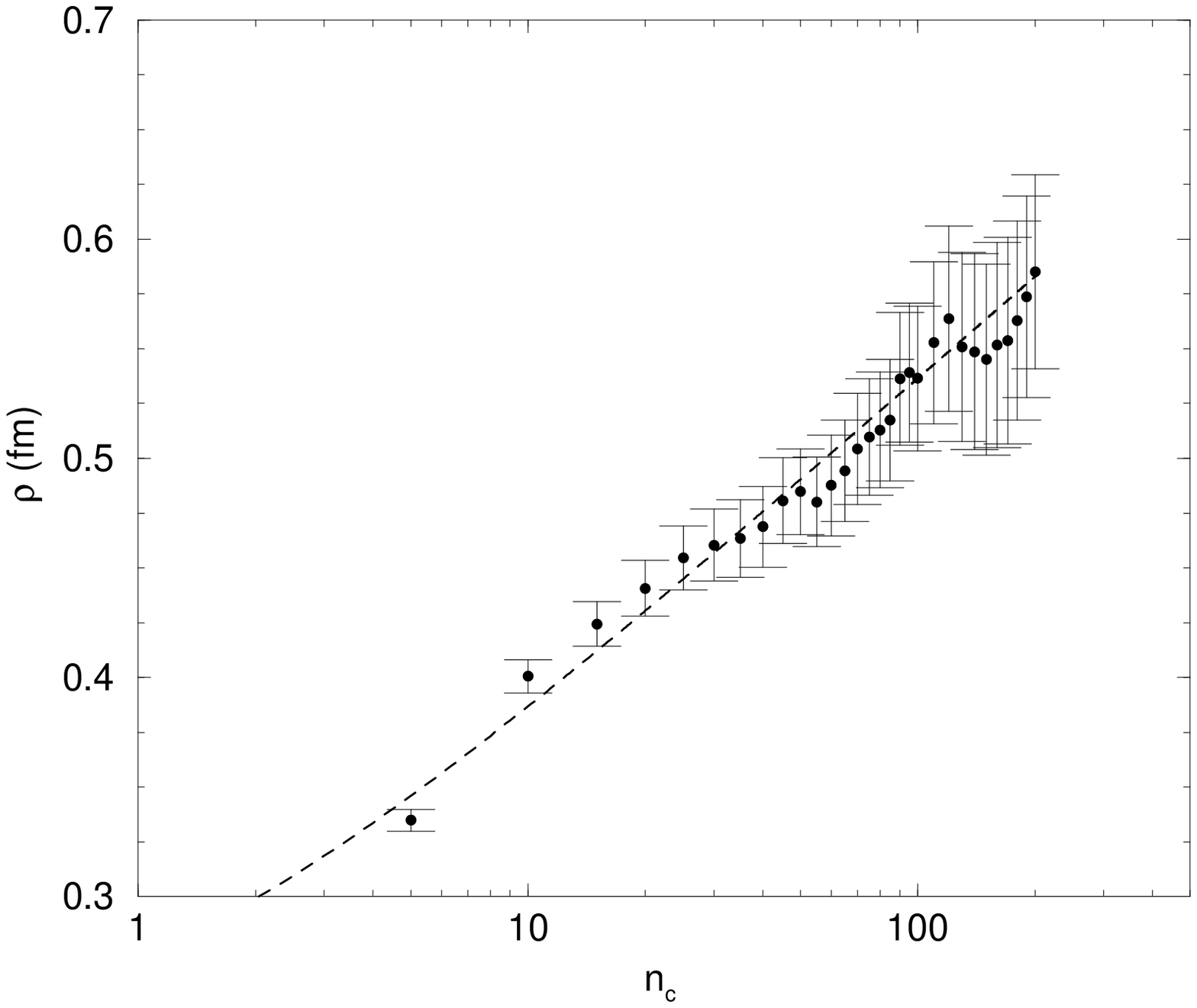} \\
\vspace*{-0.5cm}
\end{tabular}
\caption{\small {\it Results of the combined fit for the instanton density and radius 
as a function of the number of cooling sweeps for a $24^4$ lattice at $\beta$=6.0.}}
\vspace*{-0.5cm}
\label{extr}
\end{center}
\end{figure}

\noindent If we assumed $\rho(n_c)=cte$, the solution of Eq. (\ref{nnc}) would give 
$N(n_c)\sim\frac{N(0)}{1+\kappa n_c}$, the expression used in \cite{instanton}, 
as a first order approach. But our cooling procedure modifies instanton's size 
(See figure \ref{extr}), in a way that we phenomenologically parametrize as:

\beq
\rho(n_c)\ =\ \rho(0) (1 + a \ln( 1 + n_c ) )\ .
\label{rnc}
\eeq

\noindent We will include (\ref{rnc}) in equation (\ref{nnc}), with $\rho(0)$ the extrapolated 
radius at the thermalised situation and $a$ a constant to determine.

After performing a combined fit of our lattice results to the expressions 
(\ref{rnc}) and the one coming from the integration of (\ref{nnc}), we can fix
the initial values of the density, $n(0)$ and the radius $\rho(0)$, and the 
two constants that govern the evolution, $\lambda$ and $a$. An indirect measure of
the $\VEV{A^2}_{Inst}$ condensate is obtained.

Nevertheless, the  result of the extrapolation is highly dependent on the value 
of $\rho(0)$, which due to its logarithmic behaviour is hardly reliable. We 
therefore prefer the value of $\VEV{A^2_{Ins}}$ at the maximum of figure \ref{extr} 
as a crude estimation of the instantonic contribution to $\VEV{A^2}$.

Having a slight dependence on the lattice spacing, the result given here for 
this quantity has to be understood as a first estimation, that would require for a 
further analysis \cite{addendum}.

\section{Conclussion.}
The result of the combined fit gives a value of the instantonic  contribution
to $\VEV{A^2_{Ins}}\sim 0.4 \gev^2$, however, as  reasoned before, we will
give the value at the maximum as a more reliable, although crude, estimation
of the instantonic contribution to the $\VEV{A^2}$ condensate. So our final
result will be $\VEV{A^2_{Ins}}=1.10(10)\ \gev^2$ for the simulation at
lattice spacing $0.101\ fm$, and $1.37(16)\ \gev^2$  for the one at $0.074\
fm$.

This semiclassical evaluation of $\VEV{A^2}$, which does not run with the scale,
is difficult to relate to that appearing in the O.P.E. expansion, which does 
depend on the renormalisation scheme and scale. The typical scale of instantons 
is $\rho^{-1}\sim 0.7\gev$. Unluckly it is not possible to run the $\VEV{A^2_{O.P.E.}}$
to such a low energy, where pertubative QCD is not valid. The lowest reacheble 
energy scale is $2.6\gev$ \cite{ope,instanton}, where the $\VEV{A^2}$ condensate takes 
the value;

\beq
\VEV{A^2_{O.P.E.}(2.6\gev)}=1.4(3)(3)\gev^2 ,
\eeq

\noindent the first error coming from the OPE determination of 
the condensate renormalised at 10 $\gev$, and the second from 
higher orders in the running.

Keeping in mind the level of uncertainty of these calculations, 
we can nevertheless claim a rather encouraging agreement between
the instantonic contribution to the condensate and the one
computed from the running of the Green Functions.

\section*{Acknowledgements.}

F. de Soto wants to thank L.P.T. for its warm hospitality and Fundaci\'on
C\'amara and the MCYT (Contract BFM2002-03315) for financial support.


\begin{thebibliography}{9}

\bibitem{alat}  C.T.H. Davies {\it et al.}, \plb{345}{1992}{42}, \prd{56}{1997}{2755}, 
		G.S. Bali and K. Schilling \prd{47}{1993}{661}. 


\bibitem{alpha} S. Capitani  {\it et al.},
	Nucl. Phys. Procc. Suppl. {\bf 63} (1998) 153;
	\npb{544}{1999}{669}.


\bibitem{alles} B. Alles {\it et al.}, \npb{502}{1997}{325}


\bibitem{power} Ph. Boucaud {\it et al.} \jhep{10}{1998}{017};
		D. Becirevic {\it et al.} \prd{60}{1999}{094509}. 
		
		
\bibitem{ope}
	Ph.~Boucaud {\it  et al.} \plb{493}{2000}{315}, F. De Soto and 
	J. Rodr\'{\i}guez-Quintero \prd{64}{2001}{114003}.

\bibitem{Kondo}
	K.~I.~Kondo, {\it  et al.} prd{65}{2000}{085034}, 
        F.V. Gubarev, V.I. Zakharov, \plb{501}{2001}{28}.

\bibitem{thooft}
	G.~'t Hooft, \prd{14}{1976}{3432}, 
	[Erratum-ibid.\ \prd{18}{1976}{2199}].

\bibitem{shuryak}
	T.~Schafer and E.~V.~Shuryak,	Rev.\ Mod.\ Phys.\  {\bf 70} (1998) 323.

\bibitem{a2} Ph.~Boucaud {\it  et al.}, "A transparent expression of the A$^2$-Condensate's 
		renormalisation", to be published.

\bibitem{teper}
	M.~Teper, Phys.\ Lett.\ B {\bf 162}, 357 (1985),  Phys.\ Rev.\  {\bf D58} 014505 (1998).

\bibitem{negele}
J.~W.~Negele,
Nucl.\ Phys.\ Proc.\ Suppl.\  {\bf 73}, 92 (1999)
[arXiv:hep-lat/9810053].

\bibitem{garciaperez}
	M.~Garcia Perez, O.~Philipsen and I.~O.~Stamatescu, Nucl.\ Phys.\ B {\bf 551}, 293 (1999).

\bibitem{instanton}
	Ph. ~Boucaud {\it et al.}, Phys.\ Rev.\ D {\bf 66} (2002) 034504.

\bibitem{addendum} Addendum to Phys.\ Rev.\ D {\bf 66} (2002) 034504, in preparation.

\end{thebibliography}
\end{document}